\documentclass[onecolumn,showpacs,aps,prd]{revtex4}
\def\GammaBs{\Gamma_s}

\def\ETA{\zeta}
\usepackage{graphicx}
\usepackage{dcolumn}
\usepackage{amsmath}
\usepackage{epsfig}
\RequirePackage{xspace}

\begin{document}

%
%
% ....notation shift....
%
\def\x{x_s}
\def\y{y_s}
\def\C_y{C_{hy}}
\def\S_y{S_{hy}}
%
% ....notation shift....
%

\def\acpt{\alpha}
\def\op{{\cal O}}
\def\lsim{\mathrel{\lower4pt\hbox{$\sim$}}\hskip-12pt\raise1.6pt\hbox{$<$}\;}
\def\Dd{\psi}
\def\pp{\lambda}
\def\ket{\rangle}
\def\BAR{\bar}
\def\xba{\bar}
\def\fa{{\cal A}}
\def\fm{{\cal M}}
\def\fl{{\cal L}}
\def\ufs{\Upsilon(5S)}
\def\gsim{\mathrel{\lower4pt\hbox{$\sim$}}
\hskip-10pt\raise1.6pt\hbox{$>$}\;}
\def\ufour{\Upsilon(4S)}
\def\xcp{X_{CP}}
\def\ynotcp{Y}
\vspace*{-.5in}
\def\ETAp{\ETA^\prime}
\def\bfb{{\bf B}}
\def\fd{r_D}
\def\fb{r_B}
\def\ed{\ETA_D}
\def\eb{\ETA_B}
\def\hatA{\hat A}
\def\hatfd{{\hat r}_D}
\def\hated{{\hat\ETA}_D}
\def\D{{\bf D}}
\def\pcc{(+ charge conjugate)}

\def\uglu{\hskip 0pt plus 1fil
minus 1fil} \def\uglux{\hskip 0pt plus .75fil minus .75fil}

\def\slashed#1{\setbox200=\hbox{$ #1 $}
    \hbox{\box200 \hskip -\wd200 \hbox to \wd200 {\uglu $/$ \uglux}}}

\def\slpar{\slashed\partial}
\def\sla{\slashed a}
\def\slb{\slashed b}
\def\slc{\slashed c}
\def\sld{\slashed d}
\def\sle{\slashed e}
\def\slf{\slashed f}
\def\slg{\slashed g}
\def\slh{\slashed h}
\def\sli{\slashed i}
\def\slj{\slashed j}
\def\slk{\slashed k}
\def\sll{\slashed l}
\def\slm{\slashed m}
\def\sln{\slashed n}
\def\slo{\slashed o}
\def\slp{\slashed p}
\def\slq{\slashed q}
\def\slr{\slashed r}
\def\sls{\slashed s}
\def\slt{\slashed t}
\def\slu{\slashed u}
\def\slv{\slashed v}
\def\slw{\slashed w}
\def\slx{\slashed x}
\def\sly{\slashed y}
\def\slz{\slashed z}
\def\slE{\slashed E}

\title{
%%%%
%  \begin{flushright}
%  {BNL-HET-xx}\\
%  \end{flushright}
%
%%%%
\vskip 10mm
\large\bf
Measuring $B_s$ Width Difference 
at the $\Upsilon(5s)$
Using Quantum Entanglement
}

\author{David Atwood}
\affiliation{Dept. of Physics and Astronomy, Iowa State University, Ames,
IA 50011}
\author{Amarjit Soni}
\affiliation{ Theory Group, Brookhaven National Laboratory, Upton, NY
11973}

\begin{abstract}

About 90\% of $B_s\overline B_s$ pairs produced at the $\Upsilon(5s)$ 
resonance are initially $B_s^*\overline B_s^*$ pairs which decay 
radiatively to $B_s\overline B_s$. This implies that the $B_s\overline 
B_s$ pair will then be in an eigenstate of charge conjugation (i.e. 
$C=-1$) and therefore in an entangled state. This allows for a 
determination of $\Delta \GammaBs/\GammaBs$ and the CP phase using a 
number of possible correlations between the decays of the two $B_s$ 
mesons. In particular, we consider the time integrated correlation; the 
time ordering asymmetry and the time ordering-charge asymmetry, which in 
addition to time ordering distinguishes $B_s$ from $\overline B_s$, for 
various combinations of final states. With the statistics of about 
$O(10^7-10^8)$ $\Upsilon(5s)$ events available at B factories, we find 
that the time ordering asymmetry between suitably defined hadronic and 
flavor specific (tagging) decays offers a promising method for 
determining the width difference. The corresponding time ordering-charge 
asymmetry can also bound the mixing phase. Similar observables involving 
exclusive decays are also considered. At the super B factories with 
O(50) times 
greater luminosity time ordering and time ordering-charge asymmetries 
between inclusive and exclusive modes may also provide additional bounds 
on the phases in those decays.

\end{abstract}


\maketitle

\section{Introduction}\label{introduction}

Resonance production of B-mesons at electron-positron colliders have 
proven to be an extraordinarily effective tool to study flavor 
oscillations. The B-factories to date have largely only used the 
$\Upsilon(4s)$ resonance where the only neutral B-meson produced is the 
$B_d$. To produce the more massive $B_s$ meson it is necessary to 
operate at the $\Upsilon(5s)$ peak. Indeed this mode of operation has 
already been carried out at CLEO\cite{cleo5s} and more recently at BELLE 
\cite{belle:2008sc,Drutskoy:2006dw} 
where 100fb$^{-1}$ of data, amounting to about $10^7$ $B_s\overline B_s$ pairs has been collected.

%
%...............>>>>>>some references
%

In the last few years, much progress has been made in the study of 
mixing in the $B_s$ system at D0~\cite{dms2} and CDF~\cite{dms3}; the mass difference and width 
differences have been measured and some bounds have been placed on the 
oscillation phase~\cite{cdf07, d008}. Further progress is expected in the future and also 
at the LHCb experiment~\cite{lhcb09} when it takes data. While these experiments offer 
the advantage of high event rates, they are limited in the number of 
final states which they can observe. If new physics contributes to the 
mixing or decay in the $B_s$ system, it could result in different CP 
phases in different final states so doing mixing studies in a larger 
sample of final states is desirable. 

At a B factory the detector environment is generally cleaner than at a 
hadronic machine so that a larger set of final states can be observed, 
particularly those involving neutrals in their final state. On the other 
hand the B factory has the disadvantage that the $B_s$ mesons are less 
boosted so that oscillations with frequency $\Delta m_s$ might not be 
observable. Finally, the B-factory does have the feature that the mesons 
are produced in correlated pairs. The focus of this paper will be how to 
exploit this unique feature to obtain information about $B_s$ mixing. 

%
%
%
%............................<<<<<<<<.>>>>>>>>>>>>>>>>>>>HERE
%
%
%

In this paper we will consider how the quantum correlations between the 
$B_s$ mesons allow the determination of the width difference of the 
$B_s$ system through the correlation of inclusive final states between 
the two meson decays. Further studies of the correlations with exclusive 
or semi-exclusive ({\it i.e.} an exclusive state with several quantum 
amplitudes or polarizations) final states could further give information 
about the width difference and the CP phases.

We will show that luminosity sufficient to produce $10^7-10^8$ will 
generate sufficient statistics to carry out such studies however 
systematic errors originating from the single $B_s$ branching ratios 
would need to be addressed in order to obtain precision results.

Another way to address this problem is to take advantage of the 
correlated time evolution of the entangled meson pair. We find that time 
ordering 
asymmetries between different inclusive states are particularly 
sensitive to $\Delta\GammaBs$. A related time ordering-charge asymmetry 
can 
also 
give the tangent of the mixing phase, in spite of the fact that we 
assume the $\Delta m_s$ oscillations are too rapid to be observed in 
detail at B factories. This capability will help resolve the sign 
ambiguities in the time ordering asymmetry and correlation measurements 
which 
only give the cosine of phases. This methodology can be extended to 
correlations between inclusive and semi-exclusive states to obtain 
separately the CP phase for those decays. The key advantage of using 
these asymmetries is that they are null experiments which vanish in the 
limit of no mixing and therefore are not subject to large systematic 
errors 
from input branching ratios.

Future B factories are being considered with luminosities ~50 times the 
current machines\cite{Bona:2007qt,Akeroyd:2004mj,RMP08}.
If a fraction of this luminosity is on the 
$\Upsilon(5s)$ then mixing effects in a larger set of exclusive decays 
may be probed including final states which are more sensitive to new 
physics.

This method is complementary to the methods used at hadronic 
experiments, in particular CDF and D0 where the oscillations 
proportional to $\Delta m_s$ can be resolved. Note in particular the 
existing results from D0 and CDF\cite{dms1,dms2,dms3}.

In this paper we will generally be considering observables which should 
be within the capabilities of modern B-factories. For instance a 
B-factory with a luminosities of $O(10^{34}{\rm cm}^{-2}s^{-1})$ 
operating at the $\Upsilon(5s)$ peak produces $b\bar b$ final states 
with a cross section\cite{belle:2008sc} of $(0.302\pm 0.014){\rm nb}$. 
Thus, running 1 year there will be $O(10^8)$ such events. Upgrade plans 
for KEK have as the design luminosity 50~ab$^{-1}$. If about 10\% of 
this luminosity is delivered at the $\Upsilon(5s)$ then this will 
produce $1.5\times 10^9$ $b\overline b$ events.

There are several compelling motivations to preform these measurements. 
First of all, recent theoretical progress 
\cite{Lenz:2006hd,Beneke98,Dunietz00,Badin:2007bv} suggests that a fairly robust Standard Model 
prediction of the width difference may be possible and so measurement of 
this quantity could become a good test of the Standard Model. The 
Standard Model also predicts that the mixing and decay phases in the 
$B_s$ system are small, and the same for all final states with quark 
content $c\overline c s\overline s$ since the overall phase of this 
combination is $arg(-(V_{tb}V_{ts}^*)/(V_{cb}V_{cs}^*))\approx 0.02$. 
Thus the observation of a significant phase and/or variation in the 
phase between different $c\overline c s \overline s$ final states would 
indicate new physics.

In Section~\ref{correlated_states} we discuss the The $B_s\overline B_s$ 
correlated state, oscillation and CP violation in decays are dealt with 
in Section~\ref{CPViolation}. Inclusive and exclusive final states are 
discussed in Section~\ref{InclusiveAndExclusive}, time independent and 
time dependent effects follow in 
Sections~\ref{TimeIndependentCorrelations} and \ref{TimeDepend} 
respectively. A brief conclusion is given in Section \ref{conclusion}.

%%%%>>>>here

\section{The $B_s\overline B_s$ Correlated 
State}\label{correlated_states}

At the $\Upsilon(5s)$ peak there is sufficient energy to produce the 
final states $B_s\overline B_s$, $B_s^*\overline B_s$, $B_s\overline 
B_s^*$ and $B_s^*\overline B_s^*$. The vector state decays radiatively, 
$B_s^*\to\gamma B_s$, so in all of these cases the final state consists 
of a $B_s\overline B_s$ pair and $n=0,1$ or $2$ photons. This means that 
the charge conjugation of the final $B_s\overline B_s$ system is 
directly related to the number of photons so 
radiated\cite{Atwood:2001js}. In particular, the initial $\Upsilon(5s)$ 
state is in a $C=-1$ state so if it goes directly to $B_s\overline B_s$ 
then the meson pair must also be in a $C=-1$ state. If the transition is 
through $B^*_s\overline B_s$ or $\overline B^*_s B_s$ then the final 
$B_s\overline B_s$ pair must be in a $C=+1$ state because of the 
associated photon. Likewise if the transition is through $B_s^*B_s^*$ 
then the final state consists of two photons and a $B_s\overline B_s$ 
pair so the $B_s\overline B_s$ pair is in a $C=-1$ state. Note that this 
same argument applies if the meson pair is produced through a virtual 
photon (indeed one cannot a priori separate the two channels). Current 
results\cite{belle:2008sc} indicate that about 90\% of $B_s \overline 
B_s$ pair production is through the $B_s^*B_s^*$ (while the relative 
contributions of the other two modes are less well measured) so the 
final meson pair is in the C$=-1$ state at least 90\% of the time.

If we let $\vec k$ be the 3-momentum of the $B_s$ in the center of mass 
of the $B_s\overline B_s$ pair, 
then the wave function is constrained by the fact that for a scalar 
anti-scalar pair P=C=$(-1)^{L}$ so that for the odd L, C=$-1$ case:

\begin{eqnarray}
\Psi^{C=-1}\propto 
\frac{1}{\sqrt{2}}
\left ( |B_s(\vec k)\ket |\overline B_s(-\vec k)\ket
-|\overline B_s(\vec k)\ket | B_s(-\vec k)\ket
\right ) 
\label{PairWaveFunctionCOdd}
\end{eqnarray}

while for the even L, C=$+1$ case the wave function is: 
\begin{eqnarray}
\Psi^{C=+1}\propto 
\frac{1}{\sqrt{2}}
\left ( |B_s(\vec k)\ket |\overline B_s(-\vec k)\ket
+|\overline B_s(\vec k)\ket | B_s(-\vec k)\ket
\right ) 
\label{PairWaveFunctionCEven}
\end{eqnarray}

Let us denote $a=\sigma(e^+e^-\to B_s^{(*)}\overline 
B_s^{(*)})/\sigma(e^+e^-\to b\overline b)$ at the $\Upsilon(5s)$ peak 
and let $r$, $r^*$ and $r^{**}$ refer to the fraction of this branching 
ratio which contains 0, 1 or 2 vectors respectively. Current 
experimental results \cite{cleo5s,belle:2008sc,hfag2009} give $a=19.7\pm 
2.9\%$ and $r^{**}=90.1^{+3.8}_{-4.0}\%$. $r$ and $r^*$ are as yet 
unmeasured but are constrained by the $r^{**}$ result since 
$r+r^*+r^{**}=1$. Clearly then the $B_s\overline B_s$ pair is dominantly 
in the $\Psi^{C=-1}$ state, at least 90\% of the time.

The $B_s\overline B_s$ pairs thus produced at the $\Upsilon(5s)$ in the 
C=-1 state are therefore similar to the $B^0\overline B^0$ pairs 
produced at the $\Upsilon(4s)$. However, because of the different regime 
of mixing parameters, the quantities which can be measured using this 
effect are somewhat different.

\section{Oscillation and CP Violation in $B_s$ Decay}
\label{CPViolation}

Let us now turn our attention to the time evolution of the $B_s$ mesons. 
Unlike in the $B_d$ case, 
direct measurement of the oscillations driven by the 
mass difference are probably not practical at 
a B-factory 
since\cite{hfag2009} $\Delta m_s=17.77\pm 0.1\pm 0.07 {\rm ps}^{-1}$ is 
so large, it gives rapid oscillations which are hard to resolve. In this 
paper we will instead focus on results related to the component of 
mixing driven by the width difference. Currently the width difference is 
not well measured, results from CDF\cite{cdf07} and 
D0\cite{d008} give \cite{hfag2009} 
$|\Delta\GammaBs|=0.154^{+0.054}_{-0.070}{\rm ps}^{-1}$. This is about 
20\% of the $B_s$ decay rate where 
$\tau_s=1/\GammaBs=1.472^{+0.024}_{-0.026}{\rm ps}$. As we shall show, 
this range of mixing parameters allows time integrated and time 
dependent studies to provide information concerning the width difference 
and CP violation in the $B_s$ system.

Let us first review the standard formalism for time evolution in a 
single neutral meson\cite{textbook} and then generalize this to an 
entangled pair of mesons. We denote a state $\Psi=a_\Psi |B_s\ket + 
\overline a_\Psi |\overline B_s\ket $ which is a mixture of $B_s$ and $\overline 
B_s$ 
by:

\begin{eqnarray}
\Psi=\left [ 
\begin{array}{c}
a_\Psi\\
\overline a_\Psi
\end{array}
\right ]
\end{eqnarray}

\noindent

In this basis we can write the the general mass matrix 
subject to CPT constraints:

\begin{eqnarray}
M=\left [
\begin{array}{cc}
A&-p^2\\
-q^2&A
\end{array}
\right]
\end{eqnarray}

\noindent
Here $A$, $q$ and $p$ are general complex numbers
where we take the 
convention that $Re(pq)>0$.
This matrix then has 
complex eigenvalues $\mu_1=A-pq$ and $\mu_2=A+pq$. 
We will write these eigenvalues as 
$\mu_1=m_1-\frac{i}{2}\Gamma_1$ and
$\mu_2=m_2-\frac{i}{2}\Gamma_2$ where
$m=\frac12(m_1+m_2)$ and $\GammaBs=\frac12(\Gamma_1+\Gamma_2)$. The 
eigenfunctions corresponding to these two eigenvalues are:

\begin{eqnarray}
|\Psi_1\ket
\equiv
|B_1\ket
=
\frac{1}{\sqrt{|p|^2+|q|^2}}
\left [
\begin{array}{c}
p\\
q
\end{array}
\right]
\ \ \ \ 
|\Psi_2\ket
\equiv
|B_2\ket
=
\frac{1}{\sqrt{|p|^2+|q|^2}}
\left [
\begin{array}{c}
p\\
-q
\end{array}
\right]
\end{eqnarray}

If $\Psi(t)$ is the state at time $t$, this is related to the state at 
time=0 by:

\begin{eqnarray}
\Psi(t)=U(t)\Psi(0)
\end{eqnarray}

\noindent
where the time evolution matrix $U$ satisfies 

\begin{eqnarray}
i\frac{d}{dt}U(t)=MU(t)
\end{eqnarray}

\noindent The solution to this equation is:

\begin{eqnarray}
U=
\left[
\begin{array}{cc}
g_+                  &    e^{-\frac12\eta -i\phi} g_-     \\
 e^{\frac12\eta + i\phi} g_-    & g_+
\end{array}
\right] 
\end{eqnarray}

\noindent 
where $e^{\frac12\eta+i\phi}=q/p$ and 
$g_\pm=\frac12\left(    
e^{-i\mu_1 t}
\pm
e^{-i\mu_2 t}
\right)$. 

%%>DA>  Changes related to the referees comments. 

In the $B_s$ system $|q/p|\approx 1$, the current experimental value 
is\cite{hfag2009} $|q/p|=1.0019\pm 0.0047$. This deviation from $\eta=0$ 
is undetectably small for the methods we will discuss in this paper so 
we will proceed taking the approximation that $\eta\approx 0$. This 
experimental value of $|q/p|$ is obtained from measurement of the 
semi-leptonic asymmetry,

\begin{eqnarray}
A_{SL}(B_{s,d})&=&
\frac
{
N(\overline B_{s,d}\to \ell^+ \nu_\ell X)-
N(B_{s,d}\to \ell^-\overline \nu_\ell X)
}
{
N(\overline B_{s,d}\to \ell^+ \nu_\ell X)+
N(B_{s,d}\to \ell^-\overline \nu_\ell X)
}
\end{eqnarray}

\noindent
using the relation

\begin{eqnarray}
e^{\frac12 \eta}=\left| q/p \right|&=&
\left(
\frac{1-A_{SL}}{1+A_{SL}}
\right)^{1/4}
\end{eqnarray}
 
\noindent The average value of $A_{SL}(B_s)$ used in obtaining this 
value of $|q/p|$ is $A_{SL}(B_s)=-0.0037\pm0.0094$ where the authors of 
\cite{hfag2009} have combined $\Upsilon(4s)$ data giving $A_{SL}(B_d)$ 
with Tevatron data which gives a linear combination of $A_{SL}(B_d)$ 
with $A_{SL}(B_s)$. A more resent D0 result \cite{Abazov:2010hv} which 
is not 
included in this average gives the combined asymmetry as 
$A_{SL}^b=-0.00957\pm 0.00251(stat)\pm 0.00146(syst)$ 
where

\begin{eqnarray}
A_{SL}^b&=&
\frac
{
N( b\to \ell^+ \nu_\ell X)-
N( \overline b\to \ell^-\overline \nu_\ell X)
}
{
N( b\to \ell^+ \nu_\ell X)+
N(\overline b \to \ell^-\overline \nu_\ell X)
}
.
\end{eqnarray}

\noindent 
This is a linear combination of $A_{SL}(B_d)$ and $A_{SL}(B_s)$ which 
has a 3.2 sigma discrepancy from the Standard Model prediction for this 
quantity: $A_{SL}^b=-2.3^{+0.5}_{-0.6} \times 10^{-4}$ (predicted). This 
deviation highlights another utility of $\Upsilon(5s)$ B-factories which 
should be able to directly measure $A_{SL}^b$ as well as the $B_s$ and 
$B_d$ components separately and thus clarify the comparison of theory to 
experiment in $A_{SL}$.

%%>DA>  Changes related to the referees comments. 

Let $f_i$ be a single quantum state with $A_i$ and $\overline A_i$ being 
the decay amplitudes of $B_s$ and $\overline B_s$ to $f_i$ respectively. 
Denoting ${\bf A}_i=\left [ A_i,\overline A_i\right ]$ 
we will normalize the units of amplitude so that 
the decay rate to 
$f_i$ for a given initial state $\Psi(0)$ is:

\begin{eqnarray}
\Gamma_i(t)=\left |
{\bf A}_i U(t)\Psi(0)
\right |^2
\end{eqnarray}
 
\noindent
We can rewrite this as 

\begin{eqnarray}
\Gamma_i(t)=
Tr\left [
U^\dag(t) \ R_i\ U(t)\ \rho_0
\right ]
\end{eqnarray}

\noindent 
where $\rho_0=\Psi(0)\ \Psi(0)^\dag$ and $R_i={\bf 
A}_i^\dag{\bf A}_i$ is the decay density matrix for a single quantum 
state.

Consider now, more generally a state 
$F$ consisting of several 
individual quantum states:
$F=\left\{f_i\right\}$. The decay 
rate as a function of time thus becomes:

\begin{eqnarray}
\Gamma_F(t)=
Tr\left [
U^\dag(t) \ R_F\ U(t)\ \rho_0
\right ]
\label{FTimeWidth}
\end{eqnarray}

\noindent
where $R_F=\sum_iR_i$.
In general $R_F$ 
is Hermitian and thus
can be written in the form:

\begin{eqnarray}
R_F= 
\left[
\begin{array}{cc}
u_F+v_F                & w_F e^{i\theta_F}      \\
w_F e^{-i\theta_F}     & u_F-v_F
\end{array}
\right]
\label{uvwdef} 
\end{eqnarray}

\noindent where $u_F$, $v_F$ and $w_F$ are positive real numbers and 
$u_F^2\geq v_F^2+w_F^2$. If $F$ consists of only a single quantum state
then $u_F^2=v_F^2+w_F^2$.

Let us now expand Eqn.~(\ref{FTimeWidth}) for the initial states $B_s$, 
$\overline B_s$ and $\{B_s/\overline B_s\}$ which is an incoherent 
mixture of an equal number of $B_s$ and $\overline B_s$. The 
corresponding density matrices for these initial states are:

\begin{eqnarray}
\rho(B_s)=\left [ 
\begin{array}{cc}
1&0\\0&0
\end{array}
\right ]
\ \ \ \ 
\rho(\overline B_s)=\left [ 
\begin{array}{cc}
0&0\\0&1
\end{array}
\right ]
\ \ \ \ 
\rho(\{B_s,\overline B_s\})=
\frac12
\left [ 
\begin{array}{cc}
1 &0\\0&1
\end{array}
\right ]
\end{eqnarray}

The time dependent decay rates of these initial states to $F$ 
in the limit $\eta\to 0$ (i.e. $|q/p|=1$)
are:

\begin{eqnarray}
\Gamma(B_s\to F)&=&
e^{-\GammaBs t}
\left [
u_F\C_y+v_FC_x
- w_F \left (\S_y \cos(\phi+\theta_F)-S_x\sin(\phi+\theta_F) \right )
\right ]
\nonumber\\
\Gamma(\overline B_s\to F)&=&
e^{-\GammaBs t}
\left [
u_F\C_y-v_F C_x
- w_F \left (\S_y \cos(\phi+\theta_F)+S_x\sin(\phi+\theta_F)\right)
\right ]
\nonumber\\
\Gamma(\{B_s/\overline B_s\}\to F)&=&
e^{-\GammaBs t}
\left [
u_F\C_y
- w_F \S_y \cos(\phi+\theta_F)
\right ]
\end{eqnarray}

\noindent where

\begin{eqnarray}
\x=\Delta m_s/\GammaBs
\ \ \ \ \ \ 
\y=\Delta \GammaBs/(2\GammaBs)
\end{eqnarray}

\noindent and

\begin{eqnarray}
C_x=\cos(\x \GammaBs t)\ \ \ \ 
S_x=\sin(\x \GammaBs t)\ \ \ \ 
\C_y=\cosh(\y \GammaBs t)\ \ \ \ 
\S_y=\sinh(\y \GammaBs t)
\end{eqnarray}

The branching ratio to a particular final state is the time integral of 
the above. In particular we denote the branching ratio from the initial 
state $\{B_s/\overline B_s\}$ by

\begin{eqnarray}
\hat B_F=\frac{u_F-w_F\y\cos(\phi+\theta)}{(1-\y^2)\GammaBs}
\label{BhatDef}
\end{eqnarray}

\noindent which is the average of the branching ratios 
of $B_1$ and $B_2$ to this final state.

Let us now extend the above formalism to the system of a correlated 
$B_s\overline B_s$ (i.e. produced at the $\Upsilon(5s)$) state where the 
meson 1 with momentum $\vec k$ decays to state $F_1$ and meson 2 with 
momentum $-\vec k$ decays to final state $F_2$. We can write time 
dependent decay rate as:

\begin{eqnarray}
\Gamma^{\pm}_{F_1F_2}(t_1,t_2)
=S_{F_1F_2}Tr\left [
(U^\dag(t_1)R_{F_1}U(t_1))
Z_\pm
(U^\dag(t_2)R_{F_2}U(t_2))^T
Z_\pm^\dag
\right ]
\end{eqnarray}

\noindent
where the superscript T stands for transpose, 
$S_{F_1F_2}$ is a combinatorial factor; $S_{F_1F_2}=1$ if $F_1\neq 
F_2$ 
and $S_{F_1F_2}=\frac12$ if $F_1=F_2$
and

\begin{eqnarray}
Z_\pm=
\frac{1}{\sqrt{2}}
\left[
\begin{array}{cc}
0&1\\
\pm 1& 0
\end{array}
\right] 
\end{eqnarray}

\noindent is the matrix representation of the initial wave function, 
$\Psi^{C=\pm 1}$. Here $t_1$ is the time of decay for meson $\#1$ and 
$t_2$ is the time of decay for meson $\#2$.

Expanding the above for 
a $C=-1$ initial state, the result is:

\begin{eqnarray}
\Gamma^{-}_{F_1F_2}(t_1,t_2)
&=&
2 S_{F_1F_2}e^{-\GammaBs(t_1+t_2)}
\big [
(u_1u_2-w_1w_2 C^\phi_1C^\phi_2)\C_y^-
+(u_1w_2C^\phi_2-u_2w_1C^\phi_1)\S_y^-
\nonumber\\
&&
+(v_2w_1S^\phi_1-v_1w_2S^\phi_2)S^-_x
-(v_1v_2+w_1w_2 S^\phi_1S^\phi_2)C^-_x
\big ]
\label{TimeDepMinus}
\end{eqnarray}

\noindent
while for an initial $C=+1$ state, the decay rate is:

\begin{eqnarray}
\Gamma^{+}_{F_1F_2}(t_1,t_2)
&=&
2S_{F_1F_2} e^{-\GammaBs(t_1+t_2)}
\big [
(u_1u_2+w_1w_2 C^\phi_1C^\phi_2)\C_y^+
-(u_2w_1C^\phi_1+u_1w_2C^\phi_2)\S_y^+
\nonumber\\
&&
+(v_2w_1S^\phi_1+v_1w_2S^\phi_2)S^+_x
-(v_1v_2-w_1w_2 S^\phi_1S^\phi_2)C^+_x
\big ]
\label{TimeDepPlus}
\end{eqnarray}

\noindent where

\begin{eqnarray}
C^\phi_i=\cos(\phi+\theta_i)
&\ \ \ & 
S^\phi_i=\sin(\phi+\theta_i)
\nonumber\\
C^\pm_x=\cos((t_1\pm t_2)\x\GammaBs)
&\ \ \ & 
S^\pm_x=\sin((t_1\pm t_2)\x\GammaBs)
\nonumber\\ 
\C_y^\pm=\cosh((t_1\pm t_2)\y\GammaBs)
&\ \ \ & 
\S_y^\pm=\sinh((t_1\pm t_2)\y\GammaBs)
\end{eqnarray}

Integrating these results over $t_1$ and $t_2$ we obtain the correlated 
branching ratios

\begin{eqnarray}
B^{-}(F_1F_2)
&=&
\frac{2 S_{F_1F_2}}{\GammaBs^2}
\left [
(u_1u_2-w_1w_2 C^\phi_1C^\phi_2)\frac{1}{1-\y^2}
-(v_1v_2+w_1w_2 S^\phi_1S^\phi_2)\frac{1}{1+\x^2}
\right ]
\label{TimeIntMinus}
\end{eqnarray}

\begin{eqnarray}
B^{+}(F_1F_2)
&=&
\frac{2 S_{F_1F_2}}{\GammaBs^2}
\big [
(u_1u_2+w_1w_2 C^\phi_1C^\phi_2)\frac{1+\y^2}{(1-\y^2)^2}
-(u_2w_1C^\phi_1+u_1w_2C^\phi_2)\frac{2\y}{(1-\y^2)^2}
\nonumber\\
&&
+(v_2w_1S^\phi_1+v_1w_2S^\phi_2)\frac{2\x}{(1+\x^2)^2}
+(v_1v_2-w_1w_2 S^\phi_1S^\phi_2)\frac{1-\x^2}{(1+\x^2)^2}
\big ]
\label{TimeIntPlus}
\end{eqnarray}

\noindent
where these are the ratios with respect to the total number of 
$B_s\overline B_s$ pairs produced in each of the two CP states. 
The fraction of $b\overline b$ events at the $\Upsilon(5s)$ peak 
will therefore be:

\begin{eqnarray}
B_{5s}(F_1F_2)=a\left(
(1-r^*)B^{-}(F_1F_2)
+
r^*B^{+}(F_1F_2)
\right)
\label{Combo5sBR}
\end{eqnarray}

\noindent
where we assume that no attempt is made to distinguish between the 
$C=+1$ and $C=-1$ $B_s\overline B_s$ pairs. If these cases can be 
distinguished, for instance by counting the number of photons associated 
with the system, then an improvement in the statistics may be obtained 
although such a potential improvement is somewhat limited by the fact 
that at least 90\% of the meson pairs are in the $C=-1$ state.

%
% >>>>>>>.........>>>>>>>>>>here
%

\section{Inclusive and Exclusive Final States}
\label{InclusiveAndExclusive}

The key to obtaining basic physics parameters from the correlations and 
asymmetries we will discuss below is to choose final states where there 
is some a priori knowledge of the mixing strength $w_F$ defined in 
Eqn.~(\ref{uvwdef}). To this end 
we 
will consider two opposite limits in which that is the case. On the one 
hand we will consider inclusive states which means a large fraction of 
the $B_s$ decay modes. Depending on how you select such modes, $w_F$ can 
be either 0 or related directly to $y$. On the other extreme CP 
eigenstates or related exclusive states which provide a case where 
$|w_F/u_F|=1$.

Here we will consider two categories of inclusive final states, first of 
all flavor specific ``taggable'' final states which exclude quark 
content $c \overline c s \overline s$. Second of all ``hadronic'' final 
states which include $c \overline c s \overline s$.

The set of taggable decays includes all decays where the flavor of the 
meson can be determined from the decay products. For example in the 
decay $B_s\to \mu^+\nu D_s^-$ it is known that at the instant of decay 
that there was a $B_s$ (i.e. specifically a $\overline b s$ state) and 
not a $\overline B_s$ (i.e. $b \overline s$). We will denote taggable 
decays which indicate an initial state of $B_s$ by $t+$ and taggable 
decays which indicate an initial state of $\overline B_s$ by $t-$. If we 
are not concerned with the flavor of the initial state we will denote 
the state as $t=(t+)\ \cup\ (t-)$. For such decays $w_t=0$ since 
regardless of whether they tag $B=+1$ or $B=-1$ they cannot mix between 
$B_s$ and $\overline B_s$.

These modes are a significant fraction of $B_s$ decays. They consist of 
all semileptonic decays as well as most hadronic decays which do not have 
quark content $c\overline c s\overline s$. 
For instance we can include 
many hadronic 
decays containing only one charmed meson. 
It is advantageous to be able 
to include as many decays in the taggable sample as possible; we will 
assume that the tagging rate is 30\% of all $B_s$ decays.

If there is a partial rate asymmetry in taggable decays then $v_t\neq 
0$. Initially we will not generally be considering observables which are 
particularly sensitive to this kind of effect.

The category of ``hadronic'' decays, which we denote ``$h$'' may include 
all hadronic decays or, more generally a subset of hadronic decays that 
includes decays with quark content $c\overline c s\overline s$. To 
optimize the utility of this sample, it is best to include all 
$c\overline c s\overline s$ final states and as few other hadronic 
states as possible into the ``hadronic'' sample. As we will show below, 
cuts which are tight enough to reduce the hadronic sample by about 20\% 
while passing all $c\overline c s \overline s$ states greatly improve 
statistics in some cases.

For hadronic states $v_h\neq 0$ would indicate that there is a partial 
rate 
asymmetry. As with the taggable decays, the observables we discuss in 
this paper will generally not be sensitive to $v_h$ so we will take 
$v_h=0$. 
$w_h$ however will be non-zero and is, in fact, tied to the 
$B_s$ lifetime difference since the lifetime difference arises from a 
difference in the decay rate of the eigenstates to $c \overline c 
s\overline s$ states. Note that this also leads to a phase $\theta_h$. 

Exclusive flavor neutral states which consist of a single quantum 
amplitude, such as $F= D_s^+D_s^-$, allow the direct measurement of the 
quantity $y\cos(\phi+\theta_i)$. For such states 
$w_F=\sqrt{u_F^2-v_F^2}$ so assuming that the partial rate asymmetry is 
not large, $w_F\approx u_F$ is a good approximation. Thus, the matrix 
$R_F$ depends only on $u_F$ and the CP phase $\theta_F$. In the limit of 
CP conservation this is further constrained. If $F$ is a CP eigenstate 
and $CP=+1$, $u_F=w_F$ and $\theta_F=v_F=0$ while if $F$ is a $CP=-1$ 
state, $u_F=w_F$, $\theta_F=\pi$ and $v_F=0$.

Other exclusive states such as $F=D_s^{+*}D_s^{-*}$ consist of multiple 
quantum states (in this case due to polarization). The parameter $w_i$ 
is therefore not constrained. We may however be able to obtain 
information about the relative contribution of different amplitudes that 
make up $F$ through the study of $\{B_s/\overline B_s\}\to F$. For 
instance in the case of $F=D_s^{+*}D_s^{-*}$ we can learn the 
contributions of the different polarization states through studies at a 
hadronic $B_s$ experiment.

More generally, a semi-exclusive state consists of a small set of 
inclusive states such as $F=\psi+X$. Most likely there is no simple way 
to determine $w_F$ or $\theta_F$. However, if you combine information 
from time ordering asymmetries and time ordering-charge asymmetries 
discussed 
below, you 
can obtain the phase of such a decay. The Standard Model implies that 
all such phases will be small so if a large phase is discovered in any 
semi-inclusive set, this could be evidence for new physics.

% Below we will discuss how we might 
% obtain information about these quantities using correlated $B_s$ 
% mesons 
% produced at the $\Upsilon(5s)$ as well as states which could provide 
% either good statistics or large $w_F/u_F$. 
%
%

%
% Below we will discuss how we might 
% obtain information about these quantities using correlated $B_s$ 
% mesons 
% produced at the $\Upsilon(5s)$ as well as states which could provide 
% either good statistics or large $w_F/u_F$. 
%

\section{Time Independent Correlations}
\label{TimeIndependentCorrelations}

Let us first consider the effect that mixing has on the time independent 
correlations between final states. If no mixing were present, the 
null hypothesis, the 
overall branching ratio to the state $F_iF_j$ would be given 
in a simple way by the product of the branching ratios for each of the 
$B_s$ meson giving

\begin{eqnarray}
B_{5s}^{null}(F_i F_j)=2aS_{F_iF_j}\hat B(F_i) \hat B(F_j)
\label{NullHypothesis}
\end{eqnarray}

With mixing present, we will have a deviation 
of the measured value of $B_{5s}$
from this expectation. The magnitude of this deviation will thus tell us 
about the mixing process.

In order to carry out this program, however you need to have an accurate 
value for the basic null hypothesis so $\hat B(F_i)$ and $\hat B(F_j)$ 
must be well determined. The systematic error in $B_{5s}^{null}(F_iF_j)$ 
is therefore likely to be the main limitation in using this technique to 
probe mixing.

Let us first consider the application to inclusive states and so  
apply
Eqns.~(\ref{BhatDef}, \ref{TimeIntPlus} and 
\ref{TimeIntMinus}) to $t$ and $h$ states.

In the case of taggable states, since $w_t=0$, 

\begin{eqnarray}
\hat B_t=\frac{u_t}{\GammaBs(1-\y^2)}
\label{hatBt}
\end{eqnarray} 

\noindent
The 
origin of 
$\Delta \GammaBs$ 
is the rate difference within the hadronic decays so that $y$ is related 
to the hadronic decay mixing parameters by:

\begin{eqnarray}
\y\GammaBs=w_h\cos(\phi+\theta_h)
\end{eqnarray}

\noindent
it follows then that

\begin{eqnarray}
\hat B_h=\frac{u_h-w_h\y\cos(\phi+\theta_h)}{\GammaBs(1-\y^2)}
\label{hatBh}
\end{eqnarray}

It is convenient to write the correlated branching ratios strictly in 
terms of $\hat B$ since these are separately determined experimental 
quantities. So 
turning now to the correlated branching ratios for
$tt$, $th$ and $hh$ 
final states we obtain.

\begin{eqnarray}
B_{5s}( tt)&=&a\hat B_{t}^2(1
-
(1-2r^*)\y^2)
+O(\x^{-2})
\nonumber\\
B_{5s}(hh)&=&a\hat B_{h}^2(1
-
(1-2r^*)\y^2(\hat B_h^{-1}-1)^2)       
)
+O(\x^{-2})
\nonumber\\
B_{5s}(th)&=&2a\hat B_{t}\hat B_{h}(1+(1-2r^*)\y^2
(\hat B_h^{-1}-1))
+O(\x^{-2})
\label{BulkCorrelations}
\end{eqnarray}

\noindent Here we drop the $O(\x^{-2})$ terms since $\x$ is large for 
the 
$B_s$ system.

In each of these cases, the correlated branching ratio depends only on 
$\y^2$ and other measurable branching ratios and is independent of the 
CP violating phases. In Table~\ref{ResultTable} we show the number of 
$b\overline b$ events required at the $\Upsilon(5s)$ peak to give a 
$5-\sigma$ statistical deviation from $y=0$ both in the case of $\y=0.1$ 
and $\y=0.05$. For the hadronic states, we also consider the scenarios 
where $\hat B_h=0.7$ and $\hat B_h=0.5$. From these results we see that 
applying cuts which reduce $\hat B_h$ from $0.7$ to $0.5$ will be very 
helpful in determining $\y$ with this strategy. For instance in the $hh$ 
case, $N_{5s}(5\sigma)$ is lowered from $120\times 10^6$ to $8\times 
10^6$. Note also that since $hh$, $ht$ and $tt$ correlations all measure 
$\y^2$ it makes sense to combine the results from each of these 
combinations of final states. Combining data in this way will also lead 
to a reduction in $N_{5s}(5\sigma)$.

Thus there may well be adequate statistics to carry out this program, 
either presently or in the near future 
especially since BELLE as
mentioned previously, already has accumulated appreciable 
amount of data at the $\Upsilon(5s)$. %either presently or in the near 
%future. 
However the fractional deviation of the $B_{5s}$ from 
$B_{5s}^{null}$ is $O(\y^2)\approx 1\%$. This implies that the accuracy 
of the input values for $\hat B_{t}$ and $\hat B_{h}$ needs to be less 
than 1\% for the signal to be observable.

In the limit of CP conservation where the states are CP 
eigenstates, Eqns.~(\ref{BulkCorrelations}) can be understood in terms 
of a simple argument. Consider the case of $B_{5s}(tt)$. The decay {\it 
rate} of each of the eigenstates to taggable final states is the same so 
the {\it branching ratio} will be inversely proportional to
the total decay rate. Thus

\begin{eqnarray}
B(B_1\to t)=\frac{u_t}{\GammaBs (1-\y)}
\ \ \ \ \ \ \ \ \ 
B(B_2\to t)=\frac{u_t}{\GammaBs (1+\y)}
\label{tEigenBR}
\end{eqnarray}

\noindent
so taking the average, we obtain Eqn.~(\ref{hatBt})

If the initial state is $C=-1$, then the two meson state is  
$(|B_1\ket |B_2\ket-|B_1\ket |B_2\ket)/\sqrt{2}$ 
so that 
$B^-(tt)=B(B_1\to t)B(B_2\to t)$. 
Likewise,
the $C=+1$ state is 
$(|B_1\ket |B_1\ket+|B_2\ket |B_2\ket)/\sqrt{2}$ 
so 
$B^+(tt)=\frac12 (B(B_1\to t)B(B_1\to t)+B(B_2\to t)B(B_2\to t))$. 
Putting in the eigenstate branching 
ratios
Eqn.~(\ref{tEigenBR})
into these expressions we can thus derive 
the $tt$ correlation in 
Eqn.~(\ref{BulkCorrelations}). 

To obtain the other expressions in this 
limit,  the hadronic branching ratio  
of the two CP eigenstates is

\begin{eqnarray}
B(B_1\to h)=\frac{u_h-\y\GammaBs}{\GammaBs (1-\y)}
\ \ \ \ 
B(B_2\to h)=\frac{u_h+\y\GammaBs}{\GammaBs (1+\y)}
\label{hEigenBR}
\end{eqnarray}

\noindent
which leads to the other two correlations in 
Eqn.~(\ref{BulkCorrelations}).

Let us now 
consider 
the correlations 
involving an exclusive or semi-exclusive final state 
$Y$. 
For such a final state, let us define the quantity 

\begin{eqnarray}
P_Y&=&\frac{w_Y\cos(\phi+\theta_Y)}{u_Y}
\label{PY}
\end{eqnarray}

\nonumber
\noindent which is the quantity that we wish to measure. 
%
%
%%%%%%%%%%%%%%%%%%%%%%%----->>>>>>>>>>>>>>here: edit sentence below
%
%
%
It greatly simplifies the expressions below to 
define the related quantity:

\begin{eqnarray}
\hat P_Y &=&
\frac{P_Y-\y}{1-P_Y \y}
\label{hatPY}
\end{eqnarray}

\noindent
In the limit that CP is conserved, $\hat P_Y$ is 

\begin{eqnarray}
\hat P_Y&=&
\frac{\hat B(Y^+)-\hat B(Y^-)}
{\hat B(Y)}
\end{eqnarray}

\noindent where $Y^+$ is the subset of $Y$ which is ${\rm CP}=+1$ and 
$Y^-$ is the subset of $Y$ which is ${\rm CP}=-1$. More generally, for 
taggable states, $P_t=0$ while for hadronic states $P_h=\y\hat 
B_h^{-1}/(1-\y^2)$; thus:

\begin{eqnarray}
\hat P_t=-\y
\ \ \ \ 
\hat P_h=(\hat B_h^{-1}-1)\y
\label{phath}
\end{eqnarray}

Using this notation for the branching ratio to the state $Y$:

\begin{eqnarray}
\hat B_Y
&=&
\frac{1-\y P_Y}{1-\y^2}\frac{u_Y}{\GammaBs}
\end{eqnarray}

The correlation between $Y$ and 
the $h$ and $t$ inclusive states 
are:

\begin{eqnarray}
B_{5s}(tY)&=& 2a\hat B_t\hat B_Y\left (
1 + (1-2r^*)\y\hat P_Y
\right )
+O(\x^{-2})
\\
B_{5s}(hY)&=& 2a\hat B_h\hat B_Y\left (
1 - (1-2r^*)\y(\hat B_h^{-1}-1)\hat P_Y
\right )
+O(\x^{-2})
\label{YhtCorrelations}
\end{eqnarray}

The correlation between two different states in general is:

\begin{eqnarray}
B_{5s}(Y_iY_j)&=& 2aS_{Y_iY_J} \hat B_{Y_i}\hat B_{Y_j}\left (
1 - (1-2r^*)\hat P_{Y_i} \hat P_{Y_j}
\right )
+O(\x^{-2})
\label{YYCorrelations}
\end{eqnarray}

As discussed in\cite{Atwood:2002ak,three_refs}, it is easy to understand 
these 
correlations in the limit of CP conservation. In particular, suppose 
that $Y$ is a CP=+1 eigenstate so $\hat P_Y=+1$. If we start with an 
initial charge conjugation $-1$ $B_s \overline B_s$ state, then if one 
of the mesons decays to $Y$, the other must therefore be in the $B_2$ 
state. The probability of it decaying to a taggable state is therefore 
$B(B_2\to t)$ given in Eqn.~(\ref{tEigenBR}). Conversely if we start 
with an initial charge conjugation $+1$ $B_s \overline B_s$ state, then 
if one of the mesons decays to $Y$, the other must therefore be in the 
$B_1$ state. In this case the probability of decaying to a taggable 
decay is $B(B_1\to t)$. From this we can derive the $tY$ correlation 
above in the case $\hat P_Y=1$. We can generalize this argument to a 
case where $|\hat P_Y|<1$ considering separately the $Y^+$ and $Y^-$ 
components.

In summary then the correlations between 
inclusive states in 
Eqn.~(\ref{BulkCorrelations}) 
can determine $|\y|$ while the correlations between $Y$ and $h$ or $t$ 
give $\hat P_Y$ which in turn gives $y w_Y \cos(\phi+\theta_Y)$.
If $w_Y$ can be determined  
in some way, then the cosine of the phase is determined. 

If $Y$ is flavor neutral exclusive state (i.e. a CP eigenstate), then 
$w_Y=u_Y$. For states with multiple amplitudes such as $D_s^*D_s^*$, 
$\psi\phi$, $D_sD_s\eta^\prime$, $w$ must be determined from detailed 
analysis of the final state or the state needs to be separated into its 
constituent quantum states, for instance by polarization or Dalitz plot 
analysis.
For semi-exclusive states such as $\phi+X$ there is no way with this 
kind of data to factor $P_Y$ into $w_Y$ and $\cos(\theta_Y+\phi)$, so 
more information is required to do this.

Correlations between two semi-exclusive states may also be 
used to determine $P_Y$ by using the correlation in 
Eqn.~(\ref{YYCorrelations}).

In the case where the two states are the same or have the same CP 
eigenvalue this method has the advantage that it is almost a null experiment 
and it is therefore not subject to the contamination due to systematic 
errors in the input branching ratios. Looking at 
Eqn.~(\ref{YYCorrelations}) we see that in the limit of $r^*\to 0$ and 
$\hat P_{Y_i}= \hat P_{Y_j}=\pm 1$ which would be the case if CP were 
conserved then $B_{5s}$ would be 0. This limit can be understood in 
terms of Bose statistics. In the $C=-1$ state, a $B_s$ pair consists of 
one CP=+1 (i.e. $B_1$) and one CP=-1 (i.e. $B_2$) meson so you would 
never see two decays to the same CP eigenstate.

This correlation is not directly sensitive to $y$ but it is sensitive to 
the mixing angle of the two states. As an illustration, let us consider 
the case where we have two states with $\hat B=10^{-3}$ where $r^*=0.1$ 
that are CP=+1 eigenstates. 
If there is no CP violation so 
$\hat P_{Y_i}=
\hat P_{Y_j}=+1$
then 
$B_{5s}({\rm no\ mixing})  =8\times 10^{-8}$; 
where the fact that 
it is non-zero is due to the term proportional to $r^*$.
On the other hand, suppose that there were large mixings in one or both 
of the channels so that $|\hat P_{Y_i}\hat P_{Y_j}|<<1$ and 
so $B_{5s}({\rm large\ mixing})  =4\times 10^{-7}$ about 5 times larger. 
Thus if you have $N=7.5\times 10^7$ events you can rule out the large 
mixing scenario at $5\sigma$. 

%
%
%<><><><><<><><><><><><><><><><><>here
%
%

%%DA->start

Most likely, however, one can probably not find a CP eigenstate decay 
mode with a branching ratio this high if the acceptance is factored in. 
In Table~\ref{nullcortab} we consider the number of $b\overline b$ 
required to distinguish between the minimum and maximum possible 
correlations at the $5\sigma$ level. The CP=+1 states we consider in 
particular are $D_s^+D_s^-$ where we assume that the acceptance of this 
final state is $\acpt=1\%$, and $K^+K^-$ which should have acceptance 
nearly $\acpt=100\%$. The latter state has the advantage that it has a 
significant penguin contribution and so is more likely to be influenced 
by new physics. The CP=-1 states we consider in particular are 
$\psi\eta$ where we assume that the acceptance of this final state is 
$\acpt=4\%$, and $\pi^0K_sK_s$ where we assume the acceptance is 
$\acpt=25\%$.

We can see that the numbers are large even for super B-factories. 
Perhaps the most promising cases are 
$D_s^+D_s^-/D_s^+D_s^-$
and
$D_s^+D_s^-/K^+K^-$
where 
the limiting factor is the acceptance of the $D_s^+D_s^-$.

%%DA->end

\begin{table}
\begin{tabular}{|c|c|c|c|}
\hline
Final State & Inputs 
& $N_{5s}(5\sigma)/10^6$ ($y=0.1$)
& $N_{5s}(5\sigma)/10^6$ ($y=0.05$)
\\
\hline
$\Upsilon(5s)\to tt$ & $\hat B_t=0.3$, $r^*=0.1$  & 21 
& 350
\\
$\Upsilon(5s)\to ht$ & $\hat B_h=0.7$, $\hat B_t=0.3$, $r^*=0.1$  & 25
& 410
\\
$\Upsilon(5s)\to ht$ & $\hat B_h=0.5$, $\hat B_t=0.3$, $r^*=0.1$  & 6.5
& 100
\\
$\Upsilon(5s)\to hh$ & $\hat B_h=0.7$, $r^*=0.1$  & 120
& 1900 
\\
$\Upsilon(5s)\to hh$ & $\hat B_h=0.5$, $r^*=0.1$  & 7.8
& 130
\\
$\Upsilon(5s)\to hh$; $ht$; $tt$ & $\hat B_h=0.7$, $r^*=0.1$  & 16
& 260
\\
$\Upsilon(5s)\to hh$; $ht$; $tt$ & $\hat B_h=0.5$, $r^*=0.1$  & 4.9
& 78
\\
$\Upsilon(5s)\to Yt$ & $\hat B_t=0.3$, $\hat B_Y=10^{-3}$, $P_Y=1$ , 
$r^*=0.1$  & 33
& 130
\\
$\Upsilon(5s)\to Yh$ & $\hat B_h=0.7$, $\hat B_Y=10^{-3}$, $P_Y=1$ , 
$r^*=0.1$  & 76
& 300
\\
$\Upsilon(5s)\to Yh$ & $\hat B_h=0.5$, $\hat B_Y=10^{-3}$, $P_Y=1$ , 
$r^*=0.1$  & 20
& 78
\\
%$A(ht)$ & $\hat B_h=0.7$, $\hat B_t=0.3$, $r^*=0.1$  & 1.8
%& 29
%\\
%$A(ht)$ & $\hat B_h=0.5$, $\hat B_t=0.3$, $r^*=0.1$  & 1.3
%& 21
%\\
%$A(tY)$ &  $\hat B_Y=10^{-3}$, $P_Y=1$, $\hat B_t=0.3$, $r^*=0.1$  
%& 21
%& 93
%\\
%$A(hY)$ &  $\hat B_Y=10^{-3}$, $P_Y=1$, $\hat B_h=0.7$, $r^*=0.1$  
%& 12
%& 46
%\\
\hline
\end{tabular}
\caption{
The value of $N_{5s}(5\sigma)$, the number of $b\overline b$ events at 
the $\Upsilon(5s)$ required to observe a $5\sigma$ deviation from the 
null hypothesis Eqn.~\ref{NullHypothesis}, {\it assuming perfect 
knowledge of the input branching ratios,} for the pairs of final states 
indicated. The results are shown for $\y=0.1$ and $\y=0.05$.
}
\label{ResultTable}
\end{table}

%
%
%------------------------------->>>>>>>>>>>>>here
%
%

%%DA->start

\begin{table}
\begin{tabular}{|c|c|c|c|}
\hline
Final State
&
$\hat B_i$/$\hat B_j$
&
$\acpt_i$/$\acpt_j$(\%)
&
$N_{5s}(5\sigma)$  ($10^9$)
\\
\hline

$D_s^+D_s^-$/$D_s^+D_s^-$
&
$1.1\times 10^{-2}$/$1.1\times 10^{-2}$
&
1/1
&
12
\\

$K^+K^-$/$K^+K^-$
&
$3.3\times 10^{-5}$/$3.3\times 10^{-5}$
&
100/100
&
130
\\

$D_s^+D_s^-$/$K^+K^-$
&
$1.1\times 10^{-2}$/$3.3\times 10^{-5}$
&
1/100
&
20
\\

$\psi\eta$/$\psi\eta$
&
$9\times 10^{-4}$/$9\times 10^{-4}$
&
4/4
&
110
\\

$\pi^0K_sK_s$/$\pi^0K_sK_s$
&
$3.7\times 10^{-5}$/$3.7\times 10^{-5}$
&
25/25
&
1600
\\

$\psi\eta$/$\pi^0K_sK_s$
&
$9\times 10^{-4}$/$3.7\times 10^{-5}$
&
4/25
&
210
\\

\hline
\end{tabular}
\caption{
The number of $b\overline b$ events (unlike the previous Table, here in units of $10^9$)
required to distinguish between the maximum and minimum possible 
correlation between two CP eigenstates at the $5\sigma$ level including 
the acceptances for the final states shown. 
}
\label{nullcortab}
%\ref{nullcortab}
\end{table}

\section{Time dependent Effects}
\label{TimeDepend}

Looking at 
Eqns.~(\ref{TimeDepMinus}, \ref{TimeDepPlus})
a couple of features of the time dependence are apparent. 
Let us denote $t_\pm=t_1\pm t_2$ so that
the expression for $\Gamma^-(t_1,t_2)$ is of the form 
$e^{-(\GammaBs t_+)}f(t_-)$.   
If we integrate this over $t_+$ we obtain

\begin{eqnarray}
\frac{dB^-}{dt_-}=\frac{1}{2\GammaBs} e^{-\GammaBs |t_-|} f(t_-).
\end{eqnarray}

\noindent
where $t_-$ ranges from $-\infty$ to $+\infty$ and the superscript ($-$) 
on B indicates a C-odd initial state.

The fact that $t_+$ integrates out trivially is, of course, exploited in 
the design of the $B$-factory. We can determine the function $f$ by 
binning events according to $t_-$ and taking into account the 
exponential prefactor. Because the $B$ mesons are created with proper 
motion in the lab frame, $t_-$ can be inferred by the physical separation 
between the two decay vertices. In particular, the original interaction 
vertex where the mesons are created need not be determined. This is a 
useful feature of $\Gamma^-$ since the $e^+e^-$ interaction vertex 
cannot be directly observed.

In contrast, the expression for $\Gamma^+(t_1,t_2)$ is of the form
$e^{-\GammaBs t_+}g(t_+)$.   
If we integrate out the variable $t_-$ we obtain

\begin{eqnarray}
\frac{dB^+}{dt_+}={t_+} e^{-\GammaBs t_+} g(t_+).
\end{eqnarray}

\noindent
where $t_+$ ranges from $0$ to $+\infty$.
To determine $t_+$ we do indeed need to know the location of the 
interaction vertex. It is thus more difficult to study the time 
dependence of $\Gamma^+$. In fact experimental studies of the time 
dependence will be greatly helped by the feature that the meson pair is in a $C=-1$ 
more than
$90\%$ of the time because the time difference ($t_-$) is easier to 
measure than the time sum ($t_+$) at an asymmetric B factory.

The terms proportional to $\S_y^-$ and $S_x^-$ in 
Eqn.~(\ref{TimeDepMinus}) have the property that they are antisymmetric 
under $t_1\leftrightarrow t_2$. An observable which has the same 
symmetry will therefore be sensitive to these terms. The 
$S_x^-=\sin(\Delta m_s (t_1-t_2))$ term oscillates at the rate 
$\Delta 
m_s$ so it is not readily observable at B factories by directly 
resolving the oscillations. However, the asymmetry $A^\prime_{tj}$ discussed 
below does offer the prospect of sensitivity to this kind of time 
dependence.

The simplest such observable is the time ordering asymmetry:

\begin{eqnarray}
A_{ij}=\left\{
\begin{array}{c}
+1\ {\rm if}\ t_1>t_2\cr
-1\ {\rm if}\ t_1<t_2
\end{array}
\right .
\end{eqnarray}

the expectation value of $A_{ij}$
will receive contributions from the 
coefficients of $\S_y^-$ and $S_x^-$ which we denote:

\begin{eqnarray}
Q^y_{ij}=
2(u_iw_jC^\phi_j
-u_jw_iC^\phi_i)
=
2\GammaBs^2 \hat B_i \hat B_j(1-y^2)(\hat P_j- \hat P_i) 
\ \ \ \ \ \ 
Q^x_{ij}=
2(v_jw_iC^\phi_i
-v_i w_j C^\phi_j)
\end{eqnarray}

\noindent
where for flavor neutral final states with $v_i=v_j=0$, 
the term proportional to $Q^x_{ij}$ vanishes.

The expectation value of $A_{ij}$ is, of course, just the asymmetry 
between the two decay modes according to which decays first. Since the 
two meson final state at $B$ factories has a proper motion in the lab, 
this asymmetry can be calculated by:

\begin{eqnarray}
<A_{ij}>
&=&
\frac{
({\rm cases\ where\ F_i\ happens\ later}) 
-
({\rm cases\ where\ F_j\ happens\ later}) 
}
{\rm all\ F_iF_j\ events}
\end{eqnarray}

\noindent
where ``happens later'' translates into ``decays further downstream''.

Evaluating this expectation value we obtain in the flavor neutral case:

\begin{eqnarray}
<A_{ij}>
=
(1-r^*)\left(
Q^y_{ij}\frac{\y}{1-\y^2}
\right)
=
(1-r^*)(\hat P_j-\hat P_i)\y
\end{eqnarray}

The number of $b\overline b$ events at the $\Upsilon(5s)$ required to 
observe this asymmetry with a $n-\sigma$ significance is 

\begin{eqnarray}
N_{5s}(n\sigma)
\approx 
\frac{n^2}{A^2}\frac{1}{2a\hat B_i\hat B_j} 
+O(\y^2)
=
\frac{n^2}
{2a(1-r^*)^2\hat B_i \hat B_j (\hat P_j-\hat P_i)^2 \y^2}
+O(\y^2)
\end{eqnarray}

Let us now specialize 
this formula to the 
inclusive states.
Since the two final states must be different to form this asymmetry, 
for a $th$ final 
state this becomes:

\begin{eqnarray}
N_{5s}(n\sigma; A_{ij}(th))
=
\frac{n^2\hat B_h}
{2a(1-r^*)^2\hat B_t \y^4}
\end{eqnarray}

\noindent
where the dependence on $\hat B_h$ in the numerator comes from the 
expression for $\hat P_h$ in Eqn.~\ref{phath}.

Likewise, if we correlate semi-inclusive state $Y$ with $t$ 
or $h$ states:

\begin{eqnarray}
N_{5s}(n\sigma;tY)
&=&
\frac{n^2}
{2a(1-r^*)^2\hat B_t \hat B_Y (\hat P_Y+\y)^2 \y^2}
\\
N_{5s}(n\sigma;hY)
&=&
\frac{n^2}
{2a(1-r^*)^2\hat B_h \hat B_Y (\hat P_Y+\y(\hat B_h^{-1}-1) )^2 \y^2}
\end{eqnarray}

% % % % 
%
%
%>>>>>>>>>>>>>>>>>>   bring in a new table here for this
%
%>>>>>>>>>>>>>>xxxxxx
%
%
In Table~\ref{AsymTable} we show the results for 
 $N_{5s}(5\sigma)$. Again we consider a generic state $Y$ with 
 branching 
 ratio $10^{-3}$ and $\hat P_Y$ near 1.
%
%
%
%%%%%%%%%%%%%%%%%%%%%%%%%%%%%
%
%

%%%%%%%%%%%%%%%%%%%%>>>>>>>>>>>>here*************

Looking at the value of $N_{5s}(5\sigma)$ for the asymmetry in the $ht$ 
final state, we see that the asymmetry $A_{ij}$ requires lower 
statistics to measure $\y$. This is largely due to the fact that the 
taggable decay is more likely for the long lived $B_2$ state while the 
hadronic decay is more likely for the short lived $B_1$ state and the 
two tendencies combine in the asymmetry since the dominant $C=-1$ state 
is always $B_1B_2$ by Bose statistics.

Furthermore, this is a null experiment, absent mixing the asymmetry will 
be 0 and so there is no large systematic error brought in due to the 
uncertainty of the input branching ratios.

Using the method of~\cite{Atwood:1991ka}
one can devise an observable which measures this term with optimal 
statistical efficiency. If we neglect terms of $O(r^*)$ and $O(\x^{-2})$ 
then the optimal observable to measure the $\S_y^-$ term is 

\begin{eqnarray}
T^-_y=\tanh(t_1-t_2)
\label{TyminusDef}
\end{eqnarray}

\noindent This time dependence is proportional to the ratio between the 
time ordering asymmetric term in Eqn.~(\ref{TimeDepMinus}) 
($\propto\S_y$) and the time ordering symmetric term ($\propto\C_y$). 
For small $y_s$ this observable offers some improvement, about factor of O(2), over the unweighted time order 
asymmetry.

Consider now the case where taggable states are correlated with some 
other flavor neutral state, either $h$ or $Y$. In this case we can 
take into account the sign of the tag and define the CP odd 
time
ordering-charge
asymmetry

\begin{eqnarray}
A^\prime_{tj}=B A_{tj}
\end{eqnarray}

\noindent where $B$ is $\pm 1$ for $t\pm$ taggable states, {\it i.e.}  
events where the tagging decay indicates a $B_s$ are weighted $+1$ while 
the events where the tagging decay indicates a $\overline B_s$ are 
weighted $-1$.

Since $v_{t+}=u_{t+}=u_{t-}=-v_{t-}$, this asymmetry is sensitive to 
$Q^x_{tj}$. Looking at the definitions of 
$Q^x_{tj}$ and 
$Q^y_{tj}$  we see that

\begin{eqnarray}
Q^x_{(t\pm)j}
&=&
\pm \tan(\phi+\theta_j)  Q^y_{(t\pm)j}
\end{eqnarray}

\noindent
therefore

\begin{eqnarray}
<A_{tj}>&=&(1-r^*)(\hat P_j+\y)\y
\\
<A^\prime_{tj}>&=&(1-r^*)(\hat P_j+\y)
\frac{(1-\y^2)\x}{1+\x^2}
\tan(\phi+\theta_j)
\label{asyratio}
\end{eqnarray}

\noindent 

If both of these 
asymmetries are measured then $\tan(\phi+\theta_j)$ can be determined 
from the ratio without having to separately measure $w_j$. In practice 
the Standard Model prediction for $\tan(\theta+\phi)\approx .02$ so this 
method will generally bound (or discover) large phases due to the 
presence of new physics.  

There are three distinct ways in which this pair of asymmetries ($A$ 
and $A^\prime$) may 
be 
used. 

\begin{enumerate}
\item
With the inclusive states, $A_{th}$ can be used to determine $|\y|$. 
$A^\prime_{th}$ can be used to find $\tan(\phi+\theta_h)$.

\item
With an exclusive state $Y$, $A_{tY}$ determines 
$\cos(\phi+\theta_Y)$ (see Eqns.~\ref{PY} and \ref{hatPY}) and then $A^\prime_{tY}$ 
separately gives $\tan(\phi+\theta_Y)$.

\item
For a semi-exclusive state such as $Y=\phi+X$ or $Y=\psi+X$ obtain 
$\tan(\phi+\theta_Y)$ from the ratio of  $A_{tY}$ and  $A^\prime_{tY}$.

\end{enumerate}

In all cases of phase measurement, a large phase or discrepancies in 
phase between different modes indicates new physics.

In Table~\ref{AsymTable} we show some sample calculations of the 
asymmetries that might be seen in various decay modes and the statistics 
required to obtain a $5\sigma$ signal for the asymmetry. For each 
combination of modes we use an assumed acceptance $\acpt$ and give 
product branching ratio (factoring in $\acpt$) $2a\acpt\hat B_i \hat 
B_j$ with respect to the total number of $b \overline b$ events at the 
$\Upsilon(5s)$ peak. We then give an estimated time ordering asymmetry 
$A_{ij}$ 
which allows us to calculate the number of events required for a 
$5\sigma$ signal.

For combinations with taggable decays, we can use Eqn.~(\ref{asyratio}) 
and
estimate the time ordering-charge asymmetry $\ A^\prime_{ij}$ assuming 
$\tan(\phi+\theta_i)=1$ and the corresponding $N_{5s}(5\sigma)$. 

We can compare these $N_{5s}(5\sigma)$ values for the two asymmetries 
with the $\sim 1\times 10^8$ which could be typical of current 
B-factories and $1.5\times 10^9$ for a 5~ab$^{-1}$ super B-factory 
(i.e. assuming a 50~ab$^{-1}$ luminosity with about 10\% of the running 
devoted to the $\Upsilon(5s)$). 

For the case of the inclusive combination $ht$ 
both asymmetries may be within the range of current B-factories. 
We have also included combinations of inclusive modes with exclusive 
modes which would likely require a super B factory.

First of all there is the exclusive mode $D_sD_s$ which, in the standard 
model, should be sensitive to the same phase as the $ht$ combination. 
We have considered it with a 10\% and a 1\% acceptance where in the 
latter case somewhat more than $10^8$ $b\overline b$ events are 
required. The $K^+K^-$ mode considered would have both tree and penguin 
contributions hence new physics in a QCD penguin could contribute there. 
The case of $\psi\phi$ has already been studied through oscillations at 
D0 and CDF. Polarization analysis is helpful in separating the CP even 
from CP odd amplitudes. In the cases where we consider inclusive and 
exclusive modes with $(K_s\pi^0)_D$, i.e. a $D_0$ which specifically 
decays to $K_s\pi^0$, the final state connects $D^0$ and $\overline D^0$ 
so in the Standard Model there is sensitivity to the CKM phase $\gamma$.
The more inclusive states $\psi+X$ and $\phi+X$ have larger branching 
ratios and the phase in those modes should agree with the overall mixing 
within the SM. A discrepancy of such phases would therefore indicate New 
Physics.

\section{Conclusion}
\label{conclusion}

In conclusion, with a sample of $O(10^7-10^8)$ $b\overline b$ events at 
the $\Upsilon(5s)$ peak there is the prospect of making a precision 
determination of $\Delta \GammaBs/\GammaBs$ through the study of $tt$, 
$hh$ and $th$ correlations. This will allow for the testing of the 
Standard Model prediction of the width difference.

Time independent correlations between various combinations taggable and 
hadronic decays have the disadvantage that there is a large systematic 
error originating from the input branching ratios. This can be remedied 
by using time dependent observables. One promising observable to use is 
the time ordering asymmetry between hadronic and taggable decays. The 
time ordering-charge asymmetry, which in addition to time ordering 
requires distinguishing $B_s$ from $\overline B_s$, also can constrain 
the mixing phase although much more statistics would be needed to 
measure the expected Standard Model value.

At super B factories with about fifty times more luminosity, it 
becomes feasible to consider time ordering and time ordering-charge 
asymmetries 
with 
exclusive states and taggable or hadronic decays. Choosing specific 
exclusive decay modes can thus target different physics issues. 

Time independent correlations between two exclusive CP eigenstates are 
not subject to the large systematic errors. For branching ratios about 
$10^{-3}$ these correlations can be sensitive to large CP-phases with B factory 
statistics.

\begin{table}
\begin{tabular}{|c|c|c|c|c|c|c|}
\hline
Final State ($F_i/F_j$)                      &
$\acpt$=Acceptance(\%)                          &
$2 a \acpt \hat B_i \hat B_j$ ($10^{-6}$)   &
$A_{ij}$  (\%)                                &
$N_{5s}(5\sigma;A)$ ($10^6$)                &
$A^\prime_{ij}$ ($\tan(\phi+\theta)=1$)  (\%)     &
$N_{5s}(5\sigma;A^\prime)$ ($10^6$)                    \\
\hline
$h/t$                                        	&
100\%                                        	&
84000                                        	&
1.3						&
1.8						&
0.49						&
12.6						\\
$D_s^+D_s^-/t$                                        	&
10 (1)                                        	&
120 (12)                                        	&
10 						&
21 (210)						&
3.78						&
150 (1500)						\\
$D_s^+D_s^-/h$                                        	&
10 (1)                                        	&
280 (28)                                        	&
8.6 						&
12 (120)						&
						&
						\\
$K^+K^-/t$                                        	&
100                                         	&
4.0                                         	&
10 						&
630						&
3.8						&
4400 						\\
$K^+K^-/h$                                        	&
100                                         	&
9.2                                         	&
8.6 						&
367 						&
						&
						\\
$\psi\phi/t$                                        	&
10                                         	&
11                                         	&
5.6 						&
725						&
2.1						&
5100 						\\
$\psi\phi/h$                                        	&
10                                         	&
26                                         	&
4.8 						&
417 						&
						&
						\\
$(K_s\pi^0)_D\phi/t$                                        	&
10                                         	&
0.37                                        	&
10 						&
6800						&
3.8						&
47000 						\\
$(K_s\pi^0)_D \phi/h$                                        	&
10                                         	&
0.84                                         	&
10 						&
3000 						&
						&
						\\
$\psi+X/t$                                        	&
10                                         	&
140                                         	&
1 						&
1800						&
0.38						&
13000 						\\
$\psi+X/h$                                        	&
10                                         	&
325                                         	&
1 						&
770 						&
						&
						\\
$\phi+X/t$                                        	&
50                                         	&
1700                                        	&
1 						&
150						&
0.38						&
1000 						\\
$\phi+X/h$                                        	&
50                                         	&
3900                                         	&
1 						&
63 						&
						&
						\\
$(K_s\pi^0)_D+X/t$                                        	&
10                                         	&
923                                        	&
1 						&
271						&
0.38						&
1900 						\\
$(K_s\pi^0)_D+X/h$                                        	&
10                                         	&
2200                                         	&
1 						&
116 						&
						&
						\\
$\phi\gamma/t$                                        	&
49                                         	&
2.7                                        	&
10 						&
37						&
3.8						&
260 						\\
$\phi\gamma/h$                                        	&
49                                         	&
6.3                                         	&
10 						&
16 						&
						&
						\\
\hline
\end{tabular}
\caption{
This table shows an estimate of the requirement for observing the time 
ordering and time ordering-charge asymmetry of the given pairs of final 
states shown in the first column. For each final state pair, we have 
assumed the acceptance $\acpt$. Note that for the $D_s$ pairs we have 
taken a $10\%$ and a $1\%$ scenario (shown in brackets). In the third 
column we show the effective branching ratio compared to $e^+e^-\to 
b\overline b$. In the $A_{ij}$ column we show the estimated time 
ordering asymmetry assuming that $\cos(\phi+\theta)\approx 1$ as 
expected in the Standard Model. For the CP eigenstate, $D_s^+D_s^-$ and 
$K^+K^-$ we assume that $|P|=1$, i.e. no large direct CP violation in 
the decay. For the inclusive states $\psi+X$ and $\phi+X$ we suppose the 
value of $\hat P\approx 10\%$. The next column shows $N_{5s}(5\sigma)$, 
the number of $b\overline b$ events needed to observe the given time 
ordering asymmetry with $5\sigma$ statistics. For the cases where a 
decay is with a tagged decay, we give the time ordering-charge asymmetry 
$\hat A_{ij}$ assuming that the total phase is $\tan(\phi+\theta)=1$ as 
might be the case in New Physics. 
Likewise, we also give the number of $b\overline b$ events needed to 
observe this time ordering-charge asymmetry with $5\sigma$ significance.
}
\label{AsymTable}
\end{table}

%
%>>>>>>>>>>>>>>>>><<<<<<<<<<<<here

\section*{Acknowledgements}

The work of D.~A. and A.~S. are supported in part by US DOE grant Nos.
DE-FG02-94ER40817 (ISU) and DE-AC02-98CH10886 (BNL).


\begin{thebibliography}{99}
\bibitem{AmbiguityNote}
Ultimately there is thus a four fold ambiguity under the symmetries 
$\phi+\theta_i\to -(\phi+\theta_i)$ and $y\to -y;\ \phi+\theta_i\to 
\pi+\phi+\theta_i$.
\end{thebibliography}
\end{document}